# Analysis and Comparison of Different Fuzzy Inference Systems used in Decision Making for Secondary Users in Cognitive Radio Network

Ashish Upadhyay[1], Shashank Kotyan[1], Shrivishal Tripathi[1], Sandeep Yadav[2]

[1]Dr. Shyama Prasad Mukherjee International Institute of Information Technology Naya Raipur

[2]Indian Institute of Technology, Jodhpur, Rajasthan

*Abstract*- **Spectrum scarcity is one of the major challenges that the modern communication engineers are going through because of inefficient utilization of allocated frequency spectrum. The spectrum scarcity is a problem because there is not enough wavelengths/frequency to match the number of channels which are required to broadcast in a given bandwidth. Therefore, the utilization of available allocated spectrum when licensed users are not in use offers an opportunity as well as challenge, also, to increase the efficiency of spectrum utilization. Cognitive Radio offers a promising solution by reutilisation of unused allocated frequency spectrum. It helps to fulfil the demand of frequency requirement for modern communication system to accommodate more data transmission. In this optimum utilization of reuse of frequency spectrum required optimising algorithms in all parts of Cognitive Cycle. This paper focuses on designing a system based on fuzzy logic with a set of input and output parameters to obtain an optimised solution. A comparative analysis is also carried out among various types of membership functions of input and output on Mamdani Fuzzy Inference System and Sugeno Fuzzy Inference System. The proposed approach is applicable to design a better system model for a given set of rules.**

*Keywords-***Fuzzy Inference System, Cognitive Radio, Mamdani, Sugeno, Spectrum Scarcity.**

## I. Introduction

Cognitive radio is a relatively new field of vision for improving the utilization of expensive natural frequency resource, the radio electromagnetic spectrum [1, 2]. Cognitive Radio offers a mechanism to improve the spectrum utilisation by allotting the unutilised spectrum of licensed users (Primary User) to unlicensed users (Secondary User) such that spectral efficiency is improved [3]. The allocation of bandwidth to secondary user should be done in such a way that it doesn't hampers or degrades the broadcast of the primary user. Moreover, secondary user transmission needs to be so efficient that the moment primary user comes in to the scenario for transmission, secondary user sense the presence of primary user and vacate the spectrum without causing any interference to primary user.

This spectrum utilization of secondary user can be understood by an example of a person having multiple cars for rent, when a person who lend the cars, and it is not in use at present, then the other person who is in need may use it, but as soon as the lender requires a particular vehicle, the borrower has to return back the vehicle to the lender. The example is analogous to the spectrum allocation in cognitive radio where vehicle is analogous to the bandwidth, the lender is analogous to the primary user and the borrower is analogous to secondary user.

Another example can be of a lab where a faculty has received multiple licenses of simulation software which are installed in a computer. When a lab is scheduled multiple users who are students, lab assistants and the faculty are using the software simultaneously. In this example, faculty is analogous to primary user, students and lab assistants are analogues to secondary users and the licensed software is analogous to bandwidth and instances of the software's that are available are analogous to the available band.

## II. Related Works

In the literature, a lot of work is proposed using various models; a variety of comparative studies is carried out on various models applied in different domains of real life problems. Y. Chen and Y. Wang et al made a comparison of Fuzzy Inference Systems (FIS) for Traffic Flow Prediction and discussed in terms of complexity of model, time for executing rules or evaluating output, resistance to noise, consistency of the system, and amount of missing data [4]. A. Kaur et al compared the two models for air conditioning system and concluded that performances of the FISs are similar, but by using Sugeno-type FIS model allows the air conditioning system to work at its full capacity [5]. A. Shleeg et al designed the two FIS in the medical field in order to evaluate the risk of breast cancer and noticed that two FIS works in a similar manner, whereas out of the two FIS, Sugeno-type FIS allows the evaluation of risk to work at its full capacity with smooth operational performance [6]. Hegazy et al examined the performance of these two FIS for predicting prices of Fund and asserted that the performance of Sugeno method is better than that of Mamdani for the same fuzzy technique, and the performance of Sugeno Gaussian membership function (MF) usually provides better results in comparison of other membership function [7]. K. Jain et al described the two FIS to evaluate the gate opening percentage for water hydro-electric power plant dam reservoir and concluded that the FIS performances are quite similar, on the other hand Sugeno-type FIS allows the evaluation of gate opening work at its full capacity with smooth operational performance [8]. V. Kansal et al reached the same conclusion for the control water flow rate in a raw mill [9].

Therefore, it is evident and can be concluded that from these comparative studies that these two FIS are same in many respects, but the performance of Sugeno FIS is better than Mamdani FIS for the same fuzzy technique. However, in case of decision making techniques considering Sugeno and Mamdani FIS's less number of parameters (input/output) are considered. Therefore, in the proposed work, for decision making at the receiver end a comparative analysis of both the models, output for different inputs parameters as well as for different types of membership functions (MF) are carried out.

## III. SIMULATION MODEL AND PARAMETERS

In the proposed work MATLAB® Fuzzy Logic Toolbox™ is used for the simulation and analysis. Fuzzy Logic Toolbox™ of MATLAB® has apps, functions and a Simulink® block for analysis and simulation, and a GUI for designing Fuzzy inference systems. The MATLAB® toolbox offers two forms of configurable FIS namely Mamdani and Sugeno. The available functions in MATLAB® provide common methods for simulation and analysis such as, like fuzzy clustering and adaptive neuro-fuzzy learning. Moreover, it also facilitates to implement simple logic rules, which corresponds to complex behavioural problems, in a fuzzy inference system [10].

When the situations are encountered where results are not exact but approximate, then fuzzy logic is preferred [11]. It was first introduced by L. Zadeh [12]. Fig. 1 shows the fuzzy logic block diagram [13, 20]. The possible lists of input parameters, which are considered in the proposed work, are listed in Table I. A brief introduction the parameters, which are considered in the proposed work, are as:

1) **Signal strength or Field Strength**: This parameter is used to determine the output power of the transmitter as received by secondary users' antenna at a distance from the transmitting antenna [14].
2) **Spectrum Demand**: It is used to determine how many secondary users are competing for a given bandwidth [14].
3) **Signal to Interference Noise Ratio**: It is the ratio of desired signal from the transmitter to the undesired signal (noise) plus interference from other transmitters [14, 15].

$$SINR_{db} = 10 \log \frac{\text{desired signal}}{\text{interference signal} + \text{noise signal}}$$

4) **Interference**: Primary User interference is measured using interference temperature model. It is a metric proposed by Federal Communications Commission (FC(C) for interference analysis. It examines the maximum allowed Radio Frequency (RF) interference acceptable at a receiving antenna given by,

$$T_I(f_C, B) = \frac{P_I(f_C, B)}{k \times B}$$

where $P_I(f_c, (B)$ is the average interference power in Watts centred at $f_c$ covering bandwidth B measured in Hertz and $T_I$ is specified in Kelvin, and Boltzmann's constant k is $1.38 \times 10^{-23}$ Joules [15].

5) **Channel Quality**: Its value is based on Received Signal Strength Indication (RSSI) and Bit Error Rate (BER) [15].
6) **Susceptibility**: It defines the rate of change of channel selection due to the cause of some change in parameter meaning, when we switch to another channel its value tells that how much susceptibility in the channel is acceptable for the successful transmission to ensue and

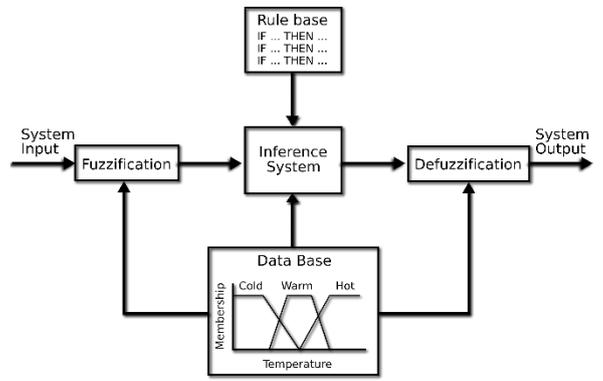

Fig. 1. Fuzzy Logic and Defuzzification

TABLE I

Representing Inputs and Outputs taken into consideration

| Input /Output | Parameter Name |
|---|---|
| Input 1 | Signal strength |
| Input 2 | Spectrum Demand |
| Input 3 | Signal To Noise Ratio (Signal to Interference plus Noise Ratio) |
| Input 4 | Interference |
| Input 5 | Channel Quality |
| Input 6 | Susceptibility |
| Input 7 | Spectrum Utilisation Efficiency |
| Input 8 | Degree of Mobility |
| Input 9 | Distance to Primary User |
| Input 10 | Secondary User Traffic Intensity |
| Input 11 | Bandwidth Allocation Traffic Intensity |
| Input 12 | Access Latency |
| Input 13 | Traffic Priority |
| Output 1 | Channel Selection Probability |
| Output 2 | Handoff Status |
| Output 3 | Channel Gain |
| Output 4 | Access the Spectrum |
| Output 5 | Access Latency |
| Output 6 | Bandwidth Allocation |

to maximize the throughput, too [15]. It is evaluated mathematically as:

$$\begin{aligned}&\text{Channel Susceptibility} \\ &= \text{Total free time of a channel} \\ &\times 100 \text{ (Total usage time of a channel} \\ &\times \text{ Total number of arrivals)} \\ &+ \text{Total free time of a channel}\end{aligned}$$

7) **Spectrum Utilisation Efficiency ($\eta_s$)**: The ratio of the spectrum band which will be allocated to the secondary user to the available band [16].

$$\eta_s = \frac{\text{Spectrum Band for Secondary User}}{\text{Available Specrum Band}}$$

8) **Degree of Mobility**: Mobility of secondary users reduces the capability of detecting signal from the primary users and if the secondary user is unable to detect the primary signal then it is possible that it will incorrectly determine that the spectrum is unused which will lead to interference to other users [16]. (Hidden node problem)
9) **Distance to Primary User**: It is the distance between primary user and secondary user. Since, the location of the primary users is unknown therefore we can consider Signal-to-Noise Ratio (SNR) as a proxy for distance as

SNR is directly proportional to distance [16] SNR at the secondary user, $\gamma_s$, is evaluated as:

$$\gamma_S = 10\log\left(\frac{\text{Transmission power of the primary user}}{\text{Noise power denoted by } \sigma_1^2}\right)$$

10) **Secondary User Traffic Intensity**: The queue storing all the entries of secondary users is referred as a Secondary Users Queue (SUQ) and the rate of entries made in SUQ is termed as secondary user traffic intensity [17].

11) **Bandwidth Allocation Traffic Intensity**: The queue which stores all entries of Secondary Users as well as Primary User is referred as a bandwidth allocation queue (BAQ) and the traffic i.e., how fast there are entries in the queue is referred as Bandwidth Allocation Traffic Intensity [17].

12) **Access Latency (T)**: Time interval between the request initiated by the Secondary User and the request granted for required bandwidth is known as access latency, and its value is evaluated as given in [17, 18].

$$T = \frac{N_1 + N_2}{\lambda_1 + \lambda_2} = \frac{\frac{\rho_1}{1-\rho_1} + \rho_2(1-P_B)}{\lambda_1 + \lambda_2}$$

13) **Traffic Priority**: It determines the nature of the traffic in the scenarios when Secondary User is ready to transmit. In addition to that, the real-time applications have higher priority than the non-real-time applications [17].

14) **Channel Selection Probability**: This parameter is determined to know the channel selection by Secondary User [14].

15) **Handoff Status**: Secondary user uses different frequency band or Handoffs when secondary users' Quality of Service (QoS) degrades [15].

16) **Channel Gain (H)**: It is defined as given in [15]:

$$H = \frac{\text{Signal received by Receiver} - \text{Noise Term}}{\text{Signal sent}}$$

17) **Access the Spectrum**: It is used to determine a Fuzzy variable which is most suitable for secondary user in having the rights to access the spectrum [16].

18) **Bandwidth Allocation**: A Factor to decide the bandwidth that may be allocated to the Secondary User [17].

## IV. RESULTS AND EXPLANATIONS

Decision process is examined by considering different possible combinations of input parameters, as mentioned in Table I, in their different possible states (Very High, High, Moderate, Low, Very Low, etc.) in the Table II to Table VII. The different possible states values of input parameters have been filled by following the rule as;

$$IF\ \left(\prod_{i=1}^{n} Input_i\ has\ fuzzy\ logic\ F_x\right)$$

$$THEN\ (Output\ has\ Fuzzy\ logic\ F_y)$$

Where $F_x$ and $F_y \in$ Fuzzy Logic

TABLE II

Represents Dependence of Channel Selection on Signal strength, Spectrum Demand and Signal to Interference and Noise Ratio

| Input 1 Signal strength | Input 2 Spectrum Demand | Input 3 Signal to Noise Ratio | Decision 1 Channel Selection Possibility |
|---|---|---|---|
| Very High | Very High | Very High | Moderate |
| Very High | Very High | High | Low |
| Very High | Very High | Moderate | Very Low |
| Very High | Very High | Low | Very Low |
| Very High | Very High | Very Low | Very Low |
| Very High | High | Very High | High |
| Very High | High | High | Moderate |
| Very High | High | Moderate | Low |
| Very High | High | Low | Very Low |
| Very High | High | Very Low | Very Low |
| Very High | Moderate | Very High | Very High |
| Very High | Moderate | High | High |
| Very High | Moderate | Moderate | Moderate |
| Very High | Moderate | Low | Low |
| Very High | Moderate | Very Low | Very Low |
| Very High | Low | Very High | Very High |
| Very High | Low | High | Very High |
| Very High | Low | Moderate | High |
| Very High | Low | Low | Moderate |
| Very High | Low | Very Low | Low |
| Very High | Very Low | Very High | Very High |
| Very High | Very Low | High | Very High |
| Very High | Very Low | Moderate | Very High |
| Very High | Very Low | Low | High |
| Very High | Very Low | Very Low | Moderate |
| High | Very High | Very High | Moderate |
| High | Very High | High | Low |
| High | Very High | Moderate | Very Low |
| High | Very High | Low | Very Low |
| High | Very High | Very Low | Very Low |
| High | High | Very High | High |
| High | High | High | Moderate |
| High | High | Moderate | Low |
| High | High | Low | Very Low |
| High | High | Very Low | Very Low |
| High | Moderate | Very High | Very High |
| High | Moderate | High | High |
| High | Moderate | Moderate | Moderate |
| High | Moderate | Low | Low |
| High | Moderate | Very Low | Very Low |
| High | Low | Very High | Very High |
| High | Low | High | Very High |
| High | Low | Moderate | High |
| High | Low | Low | Moderate |
| High | Low | Very Low | Low |
| High | Very Low | Very High | Very High |
| High | Very Low | High | Very High |
| High | Very Low | Moderate | Very High |
| High | Very Low | Low | High |
| High | Very Low | Very Low | Moderate |
| Moderate | Very High | Very High | Moderate |
| Moderate | Very High | High | Low |
| Moderate | Very High | Moderate | Low |
| Moderate | Very High | Low | Very Low |
| Moderate | Very High | Very Low | Very Low |
| Moderate | High | Very High | High |
| Moderate | High | High | Moderate |
| Moderate | High | Moderate | Low |
| Moderate | High | Low | Low |
| Moderate | High | Very Low | Very Low |
| Moderate | Moderate | Very High | Very High |
| Moderate | Moderate | High | High |
| Moderate | Moderate | Moderate | Moderate |
| Moderate | Moderate | Low | Low |
| Moderate | Moderate | Very Low | Very Low |
| Moderate | Low | Very High | Very High |
| Moderate | Low | High | Very High |
| Moderate | Low | Moderate | High |
| Moderate | Low | Low | Moderate |
| Moderate | Low | Very Low | Low |
| Moderate | Very Low | Very High | Very High |
| Moderate | Very Low | High | Very High |
| Moderate | Very Low | Moderate | Very High |
| Moderate | Very Low | Low | High |
| Moderate | Very Low | Very Low | Moderate |
| Low | Very High | Very High | Moderate |
| Low | Very High | High | Low |
| Low | Very High | Moderate | Low |
| Low | Very High | Low | Very Low |
| Low | Very High | Very Low | Very Low |
| Low | High | Very High | High |
| Low | High | High | Moderate |
| Low | High | Moderate | Low |
| Low | High | Low | Very Low |
| Low | High | Very Low | Very Low |
| Low | Moderate | Very High | Low |
| Low | Moderate | High | Low |
| Low | Moderate | Moderate | Very Low |
| Low | Moderate | Low | Low |
| Low | Moderate | Very Low | Very Low |
| Low | Low | Very High | Very High |
| Low | Low | High | Very High |
| Low | Low | Moderate | Moderate |
| Low | Low | Low | Low |
| Low | Low | Very Low | Low |
| Low | Very Low | Very High | Very High |
| Low | Very Low | High | Very High |
| Low | Very Low | Moderate | High |
| Low | Very Low | Low | Moderate |
| Low | Very Low | Very Low | Moderate |
| Very Low | Very High | Very High | Moderate |
| Very Low | Very High | High | Low |
| Very Low | Very High | Moderate | Low |
| Very Low | Very High | Low | Very Low |
| Very Low | Very High | Very Low | Very Low |
| Very Low | High | Very High | High |
| Very Low | High | High | Moderate |
| Very Low | High | Moderate | Low |
| Very Low | High | Low | Very Low |
| Very Low | High | Very Low | Very Low |
| Very Low | Moderate | Very High | Low |
| Very Low | Moderate | High | Low |
| Very Low | Moderate | Moderate | Very Low |
| Very Low | Moderate | Low | Very Low |
| Very Low | Moderate | Very Low | Very Low |
| Very Low | Low | Very High | Very High |
| Very Low | Low | High | Very High |
| Very Low | Low | Moderate | Moderate |
| Very Low | Low | Low | Low |
| Very Low | Low | Very Low | Very Low |
| Very Low | Very Low | Very High | Very High |
| Very Low | Very Low | High | Very High |
| Very Low | Very Low | Moderate | High |
| Very Low | Very Low | Low | Moderate |
| Very Low | Very Low | Very Low | Very Low |

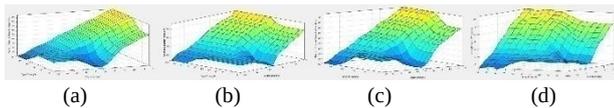

(a) (b) (c) (d)

Fig. 2 Graphical analysis of Channel selection probability *(z-axis)* against Signal strength *(x-axis)* and Spectrum demand *(y-axis)* (a) Sugeno Constant (b) Sugeno Linear (c) Mamdani Triangular (d) Mamdani Gaussian

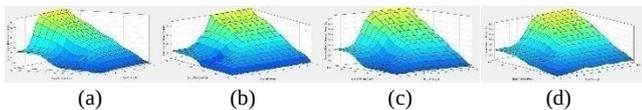

(a) (b) (c) (d)

Fig. 3 Graphical Analysis of Channel selection probability *(z-axis)* against Signal strength *(x-axis)* and SNR *(y-axis)* (a) Sugeno Constant (b) Sugeno Linear (c) Mamdani Triangular (d) Mamdani Gaussian

Fig. 2 exhibits the channel selection probability *(z-axis)* against signal strength *(x-axis)* and spectrum demand *(y-axis)*, keeping SNR constant. It is observed that channel selection probability is good when signal strength is maximum and spectrum demand is minimum. Fig. 3 presents the channel selection probability *(z-axis)* against signal strength *(x-axis)* and SNR *(y-axis)*, keeping spectrum demand constant. It can be concluded that channel selection probability for constant spectrum demand, depends more on signal strength than SNR.

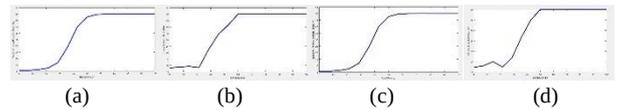

(a) (b) (c) (d)

Fig. 4 Comparative analysis of Channel selection probability *(y-axis)* against Signal strength *(x-axis)* (a) Sugeno Constant (b) Sugeno Linear (c) Mamdani Triangular (d) Mamdani Gaussian

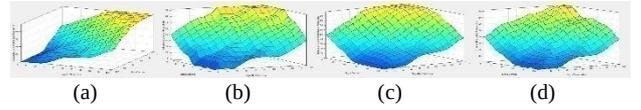

(a) (b) (c) (d)

Fig. 5 Graphical analysis of Channel selection probability *(z-axis)* against Spectrum demand *(x-axis)* and SNR *(y-axis)* (a) Sugeno Constant (b) Sugeno Linear (c) Mamdani Triangular (d) Mamdani Gaussian

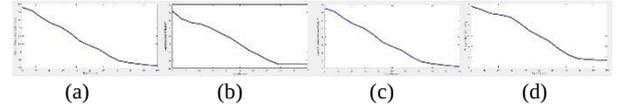

(a) (b) (c) (d)

Fig. 6 Comparative analysis of Channel selection probability *(y-axis)* against Spectrum demand *(x-axis)* (a) Sugeno Constant (b) Sugeno Linear (c) Mamdani Triangular (d) Mamdani Gaussian

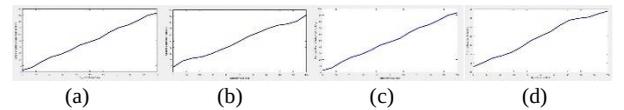

(a) (b) (c) (d)

Fig. 7 Comparative analysis of Channel selection probability *(y-axis)* against SNR *(x-axis)* (a) Sugeno Constant (b) Sugeno Linear (c) Mamdani Triangular (d) Mamdani Gaussian

TABLE III

Represents Dependence of Handoff Status on Signal to Interference and Noise Ratio and Interference

| Input 3 Signal to Noise Ratio | Input 4 Interference | Decision 2 Handoff Status |
|---|---|---|
| Very High | Very High | Off |
| Very High | High | Off |
| Very High | Moderate | On |
| Very High | Low | On |
| Very High | Very Low | On |
| High | Very High | Off |
| High | High | Off |
| High | Moderate | On |
| High | Low | On |
| High | Very Low | On |
| Moderate | Very High | Off |
| Moderate | High | Off |
| Moderate | Moderate | On |
| Moderate | Low | On |
| Moderate | Very Low | Off |
| Low | Very High | Off |
| Low | High | Off |
| Low | Moderate | Off |
| Low | Low | Off |
| Low | Very Low | Off |
| Very Low | Very High | Off |
| Very Low | High | Off |
| Very Low | Moderate | Off |
| Very Low | Low | Off |
| Very Low | Very Low | Off |

Fig. 4 unveils the channel selection probability *(y-axis)* against signal strength *(x-axis)*, keeping spectrum demand and SNR constant. In this case channel selection probability is better for higher signal strength. Fig. 5 shows the channel selection probability *(z-axis)* against spectrum demand *(x-axis)* and SNR *(y-axis)*, keeping signal strength constant. It can be concluded that channel selection probability for constant signal strength, depends more on SNR than signal strength. Fig. 6 reveals the channel selection probability *(y-axis)* against spectrum demand *(x-axis)*, keeping signal strength and SNR constant. In this case channel selection probability is better for lower spectrum demand.

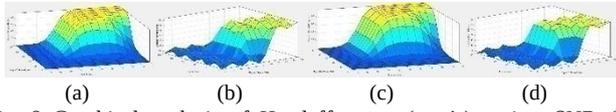
(a) (b) (c) (d)
Fig. 8 Graphical analysis of Handoff status *(z-axis)* against SNR *(x-axis)* and Interference *(y-axis)* (a) Sugeno Constant (b) Sugeno Linear (c) Mamdani Triangular (d) Mamdani Gaussian

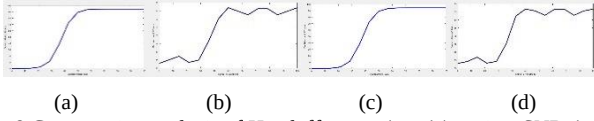
(a) (b) (c) (d)
Fig. 9 Comparative analysis of Handoff status *(y-axis)* against SNR *(x-axis)* (a) Sugeno Constant (b) Sugeno Linear (c) Mamdani Triangular (d) Mamdani Gaussian

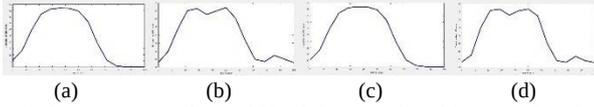
(a) (b) (c) (d)
Fig. 10 Comparative analysis of Handoff status *(y-axis)* against Interference *(x-axis)* (a) Sugeno Constant (b) Sugeno Linear (c) Mamdani Triangular (d) Mamdani Gaussian

Fig. 7 displays the channel selection probability *(y-axis)* against SNR *(x-axis)*, keeping signal strength and spectrum demand constant. It is observed that in this case that channel selection probability is improved with the increment in the SNR values. Fig. 8 exhibits the handoff status *(z-axis)* against SNR *(x-axis)* and interference *(y-axis)*. It is observed from figure that handoff status is good for higher SNR and lower interference. Fig. 9 presents the handoff status *(y-axis)* against SNR *(x-axis)*, keeping interference constant. It is observed that for constant interference handoff status is improved for higher SNR. Fig. 10 unveils the handoff status *(y-axis)* against interference *(x-axis)*, keeping signal to noise ratio constant. It is demonstrated through result that in this case handoff status is worse for too low or too high interference.

Fig. 11 exhibits the channel gain *(z-axis)* against channel quality *(x-axis)* and susceptibility *(y-axis)*. It is observed that channel gain is good for higher channel quality and lower susceptibility. Fig. 12 presents the channel gain *(y-axis)* against channel quality *(x-axis)*, keeping susceptibility constant. It is noticed that the channel gain gradually increases with increase in channel quality. Fig. 13 unveils the channel gain *(y-axis)* against susceptibility *(x-axis)*, keeping channel quality constant. It is observed that in this case of constant channel quality channel gain gradually decreases with increase in susceptibility.

Fig. 14 exhibits the spectrum accession *(z-axis)* against spectrum utilisation efficiency *(x-axis)* and degree of mobility *(y-axis)*, keeping distance to primary user constant. It is observed that accessing spectrum is easier when spectrum utilisation efficiency is large and degree of mobility is less. Fig. 15 presents the spectrum accession *(z-axis)* against spectrum utilisation efficiency *(x-axis)* and distance to primary user *(y-axis)*, keeping degree of mobility constant. It can be concluded that accessing spectrum for constant degree of mobility, depends more on efficiency of spectrum utilisation than distance to primary user. Fig. 16 unveils the spectrum accession *(y-axis)* against spectrum utilisation efficiency *(x-axis)*, keeping degree of mobility and distance to primary user constant. In this case accessing spectrum is gradually increases with efficiency in spectrum

TABLE IV

Represents Dependence of Channel Gain on Channel Quality and Susceptibility

| Input 5 Channel Quality | Input 6 Susceptibility | Decision 3 Channel Gain |
|---|---|---|
| Very High | Very High | Low |
| Very High | High | Low |
| Very High | Moderate | High |
| Very High | Low | Very High |
| Very High | Very Low | Very High |
| High | Very High | Low |
| High | High | Low |
| High | Moderate | Moderate |
| High | Low | High |
| High | Very Low | Very High |
| Low | Very High | Low |
| Low | High | Low |
| Low | Moderate | Low |
| Low | Low | Low |
| Low | Very Low | Low |
| Moderate | Very High | Low |
| Moderate | High | Low |
| Moderate | Moderate | Moderate |
| Moderate | Low | Moderate |
| Moderate | Very Low | High |
| Very Low | Very High | Low |
| Very Low | High | Low |
| Very Low | Moderate | Low |
| Very Low | Low | Low |
| Very Low | Very Low | Low |

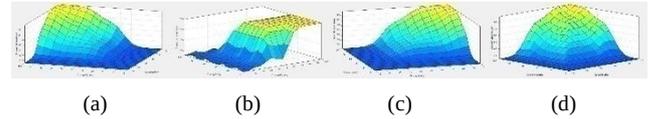
(a) (b) (c) (d)
Fig. 11 Graphical analysis of Channel gain *(z-axis)* against Channel quality *(x-axis)* and Susceptibility *(y-axis)* (a) Sugeno Constant (b) Sugeno Linear (c) Mamdani Triangular (d) Mamdani Gaussian

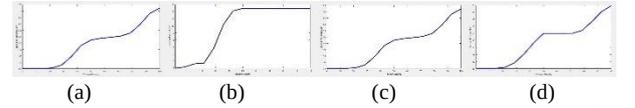
(a) (b) (c) (d)
Fig. 12 Comparative analysis of Channel gain *(y-axis)* against Channel quality *(x-axis)* (a) Sugeno Constant (b) Sugeno Linear (c) Mamdani Triangular (d) Mamdani Gaussian

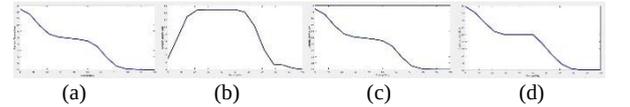
(a) (b) (c) (d)
Fig. 13 Comparative analysis of Channel gain *(y-axis)* against Susceptibility *(x-axis)* (a) Sugeno Constant (b) Sugeno Linear (c) Mamdani Triangular (d) Mamdani Gaussian

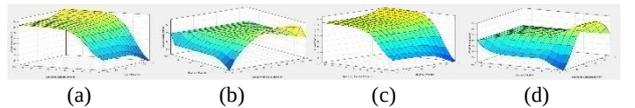
(a) (b) (c) (d)
Fig. 14 Graphical analysis of Spectrum accession *(z-axis)* against Spectrum utilisation efficiency *(x-axis)* and Degree of mobility *(y-axis)* (a) Sugeno Constant (b) Sugeno Linear (c) Mamdani Triangular (d) Mamdani Gaussian

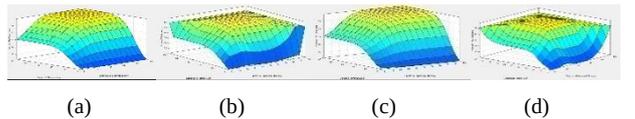
(a) (b) (c) (d)
Fig. 15 Graphical analysis of Spectrum accession *(z-axis)* against Spectrum utilisation efficiency *(x-axis)* and Distance to primary user *(y-axis)* (a) Sugeno Constant (b) Sugeno Linear (c) Mamdani Triangular (d) Mamdani Gaussian

utilisation and then becomes stagnant. Fig. 17 shows the spectrum accession *(z-axis)* against degree of mobility *(x-axis)* and distance to primary user *(y-axis)*, keeping spectrum utilisation efficiency constant. It can be concluded that accessing spectrum for constant efficiency of spectrum

TABLE V

Represents Dependence of Accessing the Spectrum on Spectrum Utilisation Efficiency, Degree of Mobility and Distance to Primary User

| Input 7 Spectrum Utilisation Efficiency | Input 8 Degree of Mobility | Input 9 Distance to Primary User | Decision 4 Access the spectrum |
|---|---|---|---|
| Small | Small | Small | Very Low |
| Small | Small | Medium | Low |
| Small | Small | Large | Low |
| Small | Medium | Small | Very Low |
| Small | Medium | Medium | Low |
| Small | Medium | Large | Moderate |
| Small | Large | Small | Low |
| Small | Large | Medium | Low |
| Small | Large | Large | Moderate |
| Medium | Small | Small | Very Low |
| Medium | Small | Medium | Moderate |
| Medium | Small | Large | High |
| Medium | Medium | Small | Very Low |
| Medium | Medium | Medium | Moderate |
| Medium | Medium | Large | High |
| Medium | Large | Small | Very Low |
| Medium | Large | Medium | Low |
| Medium | Large | Large | High |
| Large | Small | Small | Low |
| Large | Small | Medium | High |
| Large | Small | Large | Very High |
| Large | Medium | Small | Low |
| Large | Medium | Medium | High |
| Large | Medium | Large | Very High |
| Large | Large | Small | Very Low |
| Large | Large | Medium | High |
| Large | Large | Large | High |

TABLE VI

Represents Dependence of Access Latency on Secondary User Traffic Intensity and Bandwidth Allocation Traffic Intensity

| Input 10 Secondary User Traffic Intensity | Input 11 Bandwidth Allocation Traffic Intensity | Decision 5 Access Latency |
|---|---|---|
| Very Low | Absent | Very Low |
| Very Low | Present | Low |
| Low | Absent | Low |
| Low | Present | Moderate |
| Moderate | Absent | Moderate |
| Moderate | Present | High |
| High | Absent | High |
| High | Present | Very High |
| Very High | Absent | Very High |
| Very High | Present | Very High |

TABLE VII

Represents Dependence of Bandwidth Allocation on Access Latency and Traffic Priority

| Input 12 Access Latency | Input 13 Traffic Priority | Decision 6 Bandwidth Allocation |
|---|---|---|
| Very Low | Absent | Very High |
| Very Low | Present | Very High |
| Low | Absent | Moderate |
| Low | Present | High |
| Moderate | Absent | Low |
| Moderate | Present | Moderate |
| High | Absent | Low |
| High | Present | Low |
| Very High | Absent | Very Low |
| Very High | Present | Very Low |

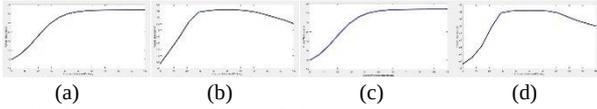

(a) (b) (c) (d)

Fig. 16 Comparative analysis of Spectrum accession *(y-axis)* against Spectrum utilisation efficiency *(x-axis)* (a) Sugeno Constant (b) Sugeno Linear (c) Mamdani Triangular (d) Mamdani Gaussian

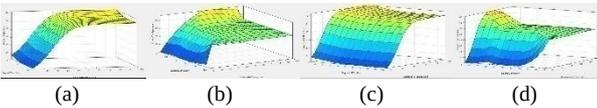

(a) (b) (c) (d)

Fig. 17 Graphical analysis of Spectrum accession *(z-axis)* against Degree of mobility *(x-axis)* and Distance to primary user *(y-axis)* (a) Sugeno Constant (b) Sugeno Linear (c) Mamdani Triangular (d) Mamdani Gaussian

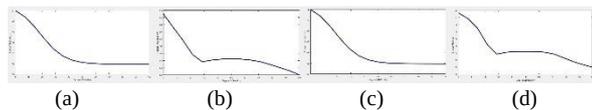

(a) (b) (c) (d)

Fig. 18 Comparative analysis of Spectrum accession *(y-axis)* against Degree of mobility *(x-axis)* (a) Sugeno Constant (b) Sugeno Linear (c) Mamdani Triangular (d) Mamdani Gaussian

utilisation, depends more on distance to primary user than degree of mobility. Fig. 18 reveals the spectrum accession *(y-axis)* against degree of mobility *(x-axis)*, keeping spectrum utilisation efficiency and distance to primary user constant. In this case accessing spectrum is gradually decreases with degree of mobility and then becomes stagnant.

Fig. 19 displays the spectrum accession *(y-axis)* against distance to primary user *(x-axis)*, keeping spectrum utilisation efficiency and degree of mobility constant. In this case accessing spectrum is gradually increases with efficiency in spectrum utilisation and then becomes stagnant. Fig. 20 exhibits the access latency *(z-axis)* against secondary user traffic intensity *(x-axis)* and bandwidth allocation traffic intensity *(y-axis)*. It is observed that latency in accessing depends highly on secondary user traffic intensity than bandwidth allocation traffic intensity.

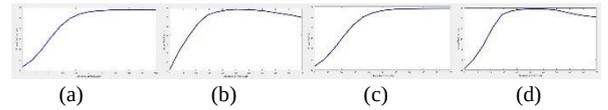

(a) (b) (c) (d)

Fig. 19 Comparative analysis of Spectrum accession *(y-axis)* against Distance to primary user *(x-axis)* (a) Sugeno Constant (b) Sugeno Linear (c) Mamdani Triangular (d) Mamdani Gaussian

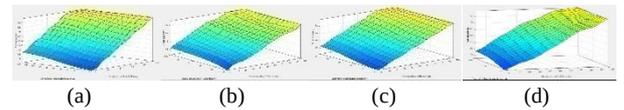

(a) (b) (c) (d)

Fig. 20 Graphical analysis of Access latency *(z-axis)* against Secondary user traffic intensity *(x-axis)* and Bandwidth allocation traffic intensity *(y-axis)* (a) Sugeno Constant (b) Sugeno Linear (c) Mamdani Triangular (d) Mamdani Gaussian

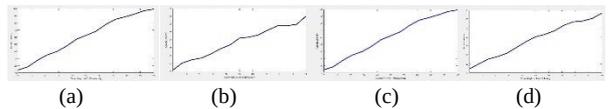

(a) (b) (c) (d)

Fig. 21 Comparative analysis of Access latency *(y-axis)* against Secondary user traffic intensity *(x-axis)* (a) Sugeno Constant (b) Sugeno Linear (c) Mamdani Triangular (d) Mamdani Gaussian

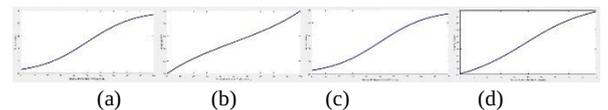

(a) (b) (c) (d)

Fig. 22 Comparative analysis of Access latency *(y-axis)* against Bandwidth allocation traffic intensity *(x-axis)* (a) Sugeno Constant (b) Sugeno Linear (c) Mamdani Triangular (d) Mamdani Gaussian

Fig. 21 presents the access latency *(y-axis)* against secondary user traffic intensity *(x-axis)*, keeping bandwidth allocation traffic intensity constant. In this case latency in accessing gradually increases with increase in intensity of secondary user traffic. Fig. 22 unveils the access latency *(y-axis)* against bandwidth allocation traffic intensity *(x-axis)*, keeping secondary user traffic intensity constant. It is noticed that latency in accessing gradually increases with increase in intensity of bandwidth allocation traffic.

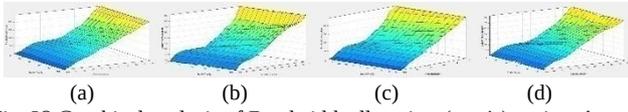

(a)　　　　(b)　　　　(c)　　　　(d)

Fig. 23 Graphical analysis of Bandwidth allocation *(z-axis)* against Access latency *(x-axis)* and Traffic priority *(y-axis)* (a) Sugeno Constant (b) Sugeno Linear (c) Mamdani Triangular (d) Mamdani Gaussian

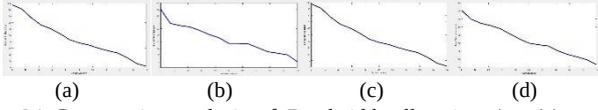

(a)　　　　(b)　　　　(c)　　　　(d)

Fig. 24 Comparative analysis of Bandwidth allocation *(y-axis)* against Access latency *(x-axis)*(a) Sugeno Constant (b) Sugeno Linear (c) Mamdani Triangular (d) Mamdani Gaussian

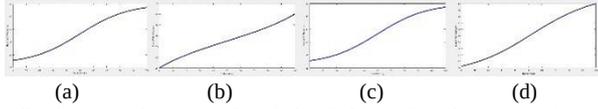

(a)　　　　(b)　　　　(c)　　　　(d)

Fig. 25 Comparative analysis of Bandwidth allocation *(y-axis)* against Traffic priority *(x-axis)* (a) Sugeno Constant (b) Sugeno Linear (c) Mamdani Triangular (d) Mamdani Gaussian

TABLE VIII

Represents Hypothetical Fuzzy Logic Ranges

| Fuzzy Logic | Range |
|---|---|
| Very Low | 0-25 |
| Low | 0-50 |
| Moderate | 25-75 |
| High | 50-100 |
| Very High | 75-100 |
| Small | 0-50 |
| Medium | 0-100 |
| Large | 50-100 |
| Absent, Off | 0-100 |
| Present, On | 0-100 |

TABLE IX

Value of channel selection possibility for four FISs when spectrum demand and SNR is kept constant and signal strength is varied

| Input 1 Signal strength | Gaussian Mamdani FIS | Linear Sugeno FIS | Triangular Mamdani FIS | Constant Sugeno FIS |
|---|---|---|---|---|
| 10 | 24.5 | 5.98 | 9.17 | 5.98 |
| 20 | 22.6 | 7.27 | 8.46 | 7.27 |
| 30 | 28.6 | 13.5 | 25.9 | 13.5 |
| 40 | 42.1 | 34.3 | 38.8 | 34.3 |
| 50 | 50 | 47.7 | 50 | 47.7 |
| 60 | 50 | 49.9 | 50 | 49.9 |
| 70 | 50 | 50 | 50 | 50 |
| 80 | 50 | 50 | 50 | 50 |
| 90 | 50 | 50 | 50 | 50 |
| 100 | 50 | 50 | 50 | 50 |

Fig. 23 exhibits the bandwidth allocation *(z-axis)* against access latency *(x-axis)* and traffic priority *(y-axis)*. It is observed that allocation of bandwidth depends highly on access latency than traffic priority. Fig. 24 presents the bandwidth allocation *(y-axis)* against access latency *(x-axis)*, keeping traffic priority constant. In this case allocation of bandwidth gradually deceases with increase in latency in accession. Fig. 25 unveils the bandwidth allocation *(y-axis)* against traffic priority *(x-axis)*, keeping access latency constant. It is observed that allocation of bandwidth gradually increases with increase in traffic priority.

Table VIII, represents the crisp output for fuzzy logic and its relation with hypothetical fuzzy logic range. For very low fuzzy logic output, the crisp output lies between 0 and 25. In case of low and small fuzzy logic output, the crisp output lies in the range of between 0 and 50. Moreover, for moderate fuzzy logic output, the crisp output lies between

TABLE X

Value of channel selection possibility for four FISs when signal strength and SNR is kept constant and spectrum demand is varied

| Input 2 Spectrum Demand | Gaussian Mamdani FIS | Linear Sugeno FIS | Triangular Mamdani FIS | Constant Sugeno FIS |
|---|---|---|---|---|
| 10 | 73.1 | 88.6 | 79.6 | 88.6 |
| 20 | 69.9 | 76.9 | 75.5 | 76.9 |
| 30 | 66.8 | 70 | 68.9 | 70 |
| 40 | 57.5 | 56.9 | 60.5 | 56.9 |
| 50 | 50 | 47.7 | 50 | 47.7 |
| 60 | 42.4 | 39.8 | 39.5 | 39.8 |
| 70 | 33 | 29.5 | 31.1 | 29.5 |
| 80 | 28.1 | 26 | 25 | 26 |
| 90 | 27.4 | 23.8 | 25 | 23.8 |
| 100 | 27.1 | 22.6 | 25 | 22.6 |

TABLE XI

Value of channel selection possibility for four FISs when signal strength and spectrum demand is kept constant and SNR is varied

| Input 3 Signal to Noise Ratio for Channel Selection | Gaussian Mamdani FIS | Linear Sugeno FIS | Triangular Mamdani FIS | Constant Sugeno FIS |
|---|---|---|---|---|
| 10 | 22.7 | 10.5 | 20.3 | 10.5 |
| 20 | 27.8 | 22.6 | 24.4 | 22.6 |
| 30 | 33 | 29.1 | 31.1 | 29.1 |
| 40 | 42.4 | 39.8 | 39.5 | 39.8 |
| 50 | 50 | 47.7 | 50 | 47.7 |
| 60 | 57.5 | 56.5 | 60.5 | 56.5 |
| 70 | 66.8 | 69 | 68.9 | 69 |
| 80 | 66.9 | 75.6 | 75.5 | 75.6 |
| 90 | 73.1 | 86.7 | 79.6 | 86.7 |
| 100 | 77.8 | 93.6 | 91.9 | 93.6 |

25 and 75. For high and large fuzzy logic output, the crisp output should lies between 50 and 100. Whereas in case of very high fuzzy logic output, the crisp output lies between 75 and 100. The medium, absent, present off and on fuzzy logic output, the crisp output values lies in the range of 0 and 100.

## V. CALCULATIONS

The Tables from IX to XXIII shows the resultant values of output parameter by keeping the other input parameters (as mentioned in Table I) constant at 50 [13]. In Table IX, spectrum demand and SNR is kept constant at 50 (moderate) and signal strength is varied from 10 to 100 with an interval of 10, and corresponding values of the four FISs are calculated. It can be observed that as signal strength increases, the value of channel selection possibility also increases and after some extent when signal strength reaches 50 channel selection possibility becomes constant at 50 (moderate) for Mamdani models and when signal strength reaches 70 it becomes constant for Sugeno models.

In Table X, signal strength and SNR is kept constant at 50 (moderate) and spectrum demand is varied in the interval of 10 from 10 to 100, and its effect are evaluated on the four FISs. It can be noticed that as spectrum demand increases, the value of channel selection possibility decreases for all the four models.

In Table XI, signal strength and spectrum demand is kept constant at 50 (moderate) and SNR is varied from 10 to 100 with a step size of 10, and its corresponding values of the four FISs are calculated. It is observed that as SNR increases, the value of channel selection possibility also increases for all the four FISs models.

TABLE XII

Value of handoff status for four FISs when interference is kept constant and SNR is varied

| Input 3 Signal to Noise Ratio for Handoff | Gaussian Mamdani FIS | Linear Sugeno FIS | Triangular Mamdani FIS | Constant Sugeno FIS |
|---|---|---|---|---|
| 10 | 36.2 | 0.0776 | 36.9 | 0.0776 |
| 20 | 33.4 | 1.61 | 34.2 | 1.61 |
| 30 | 35.2 | 14.8 | 36.6 | 14.8 |
| 40 | 56.7 | 60.1 | 55 | 60.1 |
| 50 | 66.7 | 89.2 | 67 | 89.2 |
| 60 | 63.5 | 94 | 63.1 | 94 |
| 70 | 66.3 | 94.4 | 65.8 | 94.4 |
| 80 | 66.3 | 94.4 | 65.8 | 94.4 |
| 90 | 63.5 | 94.4 | 63.1 | 94.4 |
| 100 | 66.7 | 94.4 | 67 | 94.4 |

TABLE XIII

Value of handoff status for four FISs when SNR is kept constant and interference is varied

| Input 4 Interference | Gaussian Mamdani FIS | Linear Sugeno FIS | Triangular Mamdani FIS | Constant Sugeno FIS |
|---|---|---|---|---|
| 10 | 43.3 | 38 | 45 | 38 |
| 20 | 64.8 | 80.5 | 63.4 | 80.5 |
| 30 | 66.3 | 92.9 | 65.8 | 92.9 |
| 40 | 63.5 | 94 | 63.1 | 94 |
| 50 | 66.7 | 89.2 | 67 | 89.2 |
| 60 | 56.7 | 60.1 | 55 | 60.1 |
| 70 | 35.2 | 14.8 | 36.6 | 14.8 |
| 80 | 33.4 | 1.61 | 34.2 | 1.61 |
| 90 | 36.2 | 0.0776 | 36.9 | 0.0776 |
| 100 | 33 | 0.00137 | 33 | 0.00137 |

TABLE XIV

Value of channel gain for four FISs when susceptibility is kept constant and channel quality is varied

| Input 5 Channel Quality | Gaussian Mamdani FIS | Linear Sugeno FIS | Triangular Mamdani FIS | Constant Sugeno FIS |
|---|---|---|---|---|
| 10 | 8.91 | 0.0194 | 9.02 | 0.0194 |
| 20 | 9.12 | 0.403 | 8.33 | 0.403 |
| 30 | 14.9 | 3.7 | 16 | 3.7 |
| 40 | 23.2 | 15 | 22.9 | 15 |
| 50 | 24.9 | 22.3 | 25 | 22.3 |
| 60 | 24.9 | 23.5 | 25 | 23.5 |
| 70 | 24.9 | 23.6 | 25 | 23.6 |
| 80 | 24.9 | 23.6 | 25 | 23.6 |
| 90 | 24.9 | 23.6 | 25 | 23.6 |
| 100 | 24.9 | 23.6 | 25 | 23.6 |

In Table XII, interference is kept constant at 50 (moderate) and SNR is varied from 10 to 100 with a step size of 10, and its effect is evaluated on the four FISs. It is observed that as the value of SNR increases, the value of handoff status also increases for all four models.

In Table XIII, SNR is kept constant at 50 (moderate) and the interference value is varied from 10 to 100 in the interval of 10, and its impact on the four FISs is examined. It can be observed that as SNR increases, the value of handoff status increases initially for all four models. As soon as interference value crosses 50 the value of handoff status now starts decreasing.

In Table XIV, susceptibility is kept constant at 50 (moderate) and channel quality is varied with a step size of 10 in the range of 10 to 100, and its corresponding values of the four FISs are examined. It can be noticed that with the increase in channel quality the value of channel gain increases initially for all four models. As soon as channel quality value reaches to 50 the value of channel gain becomes constant near to value of 25 (very low) for all the

TABLE XV

Value of channel gain for four FISs when channel quality is kept constant and susceptibility is varied

| Input 6 Susceptibility | Gaussian Mamdani FIS | Linear Sugeno FIS | Triangular Mamdani FIS | Constant Sugeno FIS |
|---|---|---|---|---|
| 10 | 19.7 | 9.5 | 20.3 | 9.51 |
| 20 | 24.6 | 20.1 | 24.4 | 20.1 |
| 30 | 24.9 | 23.2 | 25 | 23.2 |
| 40 | 24.9 | 23.5 | 25 | 23.5 |
| 50 | 24.9 | 22.3 | 25 | 22.3 |
| 60 | 23.2 | 15 | 22.9 | 15 |
| 70 | 14.9 | 3.7 | 16 | 3.7 |
| 80 | 9.12 | 0.403 | 8.33 | 0.403 |
| 90 | 8.91 | 0.0194 | 9.02 | 0.0194 |
| 100 | 8 | 0.000341 | 8 | 0.000341 |

TABLE XVI

Value of access the spectrum for four FISs when degree of mobility and distance to primary user is kept constant and spectrum utilisation efficiency is varied

| Input 7 Spectrum Utilisation Efficiency | Gaussian Mamdani FIS | Linear Sugeno FIS | Triangular Mamdani FIS | Constant Sugeno FIS |
|---|---|---|---|---|
| 10 | 42.2 | 44.8 | 43.3 | 44.8 |
| 20 | 48.4 | 52.8 | 49.6 | 52.8 |
| 30 | 53.9 | 60.2 | 53.8 | 60.2 |
| 40 | 54.4 | 64.4 | 54.3 | 64.4 |
| 50 | 54.4 | 66.1 | 54.5 | 66.1 |
| 60 | 54.4 | 66.7 | 54.3 | 66.7 |
| 70 | 53.9 | 66.9 | 53.8 | 66.9 |
| 80 | 52.4 | 67 | 52.7 | 67 |
| 90 | 51.1 | 67 | 51.5 | 67 |
| 100 | 50 | 67 | 50 | 67 |

TABLE XVII

Value of access the spectrum for four FISs when spectrum utilisation efficiency and distance to primary user is kept constant and degree of mobility is varied

| Input 8 Degree of Mobility | Gaussian Mamdani FIS | Linear Sugeno FIS | Triangular Mamdani FIS | Constant Sugeno FIS |
|---|---|---|---|---|
| 10 | 62.7 | 70.1 | 61.9 | 70.1 |
| 20 | 57.5 | 68.6 | 56.6 | 68.6 |
| 30 | 53.9 | 67.2 | 53.8 | 67.2 |
| 40 | 54.4 | 66.4 | 54.3 | 66.4 |
| 50 | 54.4 | 66.1 | 54.5 | 66.1 |
| 60 | 54.4 | 66 | 54.3 | 66 |
| 70 | 53.9 | 66 | 53.8 | 66 |
| 80 | 52.4 | 66 | 52.7 | 66 |
| 90 | 51.1 | 65.9 | 51.5 | 65.9 |
| 100 | 50 | 65.9 | 50 | 65.9 |

four FISs. In Table XV, channel quality is kept constant at 50 (moderate) and susceptibility is varied in the range of 10 from 10 to 100, and its impact on the four FISs is evaluated. It is observed that with the increment in the susceptibility, the value of the channel gain increases initially for all four models. Whereas, when the susceptibility value crosses to the value above 50 the value of channel gain starts decreasing. Moreover, the steepness of decrease in Sugeno models is very high as compared to that of Mamdani models.

In Table XVI, degree of mobility and distance to primary user is kept constant at 50 (moderate) and spectrum utilisation efficiency is varied from 10 to 100 in the interval of 10, and its corresponding effect on the four FISs are studied. It is perceived that as spectrum utilisation efficiency increases, the value for access the spectrum also increases for all the four FISs models.

TABLE XVIII

Value of access the spectrum for four FISs when spectrum utilisation efficiency and degree of mobility is kept constant and distance to primary user is varied

| Input 9 Distance to Primary User | Gaussian Mamdani FIS | Linear Sugeno FIS | Triangular Mamdani FIS | Constant Sugeno FIS |
|---|---|---|---|---|
| 10 | 32.5 | 23.5 | 34.6 | 23.5 |
| 20 | 43.5 | 39.4 | 45.2 | 39.4 |
| 30 | 52.3 | 54.3 | 51.8 | 54.3 |
| 40 | 54.1 | 62.7 | 53.8 | 62.7 |
| 50 | 54.4 | 66.1 | 54.5 | 66.1 |
| 60 | 54.4 | 67.4 | 54.3 | 67.4 |
| 70 | 53.9 | 67.8 | 53.8 | 67.8 |
| 80 | 52.5 | 67.9 | 52.7 | 67.9 |
| 90 | 51.3 | 68 | 51.5 | 68 |
| 100 | 50.7 | 68 | 50 | 68 |

TABLE XIX

Value of access latency for four FISs when bandwidth allocation traffic intensity is kept constant and secondary user traffic intensity is varied

| Input 10 Secondary User Traffic Intensity | Gaussian Mamdani FIS | Linear Sugeno FIS | Triangular Mamdani FIS | Constant Sugeno FIS |
|---|---|---|---|---|
| 10 | 32.4 | 22.5 | 32.9 | 22.5 |
| 20 | 37.7 | 32.5 | 36.8 | 32.5 |
| 30 | 43.2 | 42.5 | 44.1 | 42.5 |
| 40 | 52.2 | 52.5 | 41.6 | 52.5 |
| 50 | 60.3 | 62.5 | 62.5 | 62.5 |
| 60 | 64.3 | 72.5 | 64.8 | 72.5 |
| 70 | 72 | 82.5 | 71.2 | 82.5 |
| 80 | 77.4 | 90 | 78.2 | 90 |
| 90 | 79 | 95 | 78.8 | 95 |
| 100 | 86.7 | 100 | 90.5 | 100 |

In Table XVII, spectrum utilisation efficiency and distance to primary user is kept constant at 50 (moderate) and degree of mobility is tuned from 10 to 100 with a step size of 10, and its impact on the four FISs are evaluated. It can be witnessed that with the increment in degree of mobility, the value of access the spectrum decreases for all the four FISs models.

In Table XVIII, spectrum utilisation efficiency and degree of mobility is kept constant at 50 (moderate) and distance to primary user is varied from 10 to 100 in the interval of 10, and its effect is studied on the four FISs. It can be observed that as distance to primary user increases, the value of access the spectrum increases for all the four FISs models. Moreover, in case of Mamdani models, the value of access the spectrum decreases with the increase in distance to primary user, when reaching the value above 90.

In Table XIX, bandwidth allocation traffic intensity is kept constant at 50 (moderate) and secondary user traffic intensity is tuned in a step size of 10, from 10 to 100, and its effect on the four FISs is examined. It can be observed that with the increase in Bandwidth Allocation Traffic Intensity, the value of Access Latency also increases for all the four FISs models.

In Table XX, secondary user traffic intensity is kept constant at 50 (moderate) and bandwidth allocation traffic intensity is varied from 10 to 100 in the interval of 10, and corresponding to each value its subsequent effect on the four FISs are examined. It is detected that as the value of bandwidth allocation traffic intensity increases, the value of access latency also increases in case of all the four FISs models.

TABLE XX

Value of access latency for four FISs when secondary user traffic intensity is kept constant and bandwidth allocation traffic intensity is varied

| Input 11 Bandwidth Allocation Traffic Intensity | Gaussian Mamdani FIS | Linear Sugeno FIS | Triangular Mamdani FIS | Constant Sugeno FIS |
|---|---|---|---|---|
| 10 | 51.5 | 52.5 | 53.3 | 52.5 |
| 20 | 53.2 | 55 | 56 | 55 |
| 30 | 55.2 | 57.5 | 58.4 | 57.5 |
| 40 | 57.7 | 60 | 60.5 | 60 |
| 50 | 60.3 | 62.5 | 62.5 | 62.5 |
| 60 | 62.8 | 65 | 64.5 | 65 |
| 70 | 65.1 | 67.5 | 66.6 | 67.5 |
| 80 | 67 | 70 | 68.9 | 70 |
| 90 | 68.5 | 72.5 | 71.7 | 72.5 |
| 100 | 69.6 | 75 | 75 | 75 |

TABLE XXI

Value of bandwidth allocation for four FISs when traffic priority is kept constant and access latency is varied

| Input 12 Access Latency | Gaussian Mamdani FIS | Linear Sugeno FIS | Triangular Mamdani FIS | Constant Sugeno FIS |
|---|---|---|---|---|
| 10 | 66.9 | 86.3 | 66.5 | 86.3 |
| 20 | 62.3 | 67.9 | 63.2 | 67.9 |
| 30 | 56.8 | 59.2 | 55.9 | 59.2 |
| 40 | 47.8 | 36.6 | 48.4 | 46.6 |
| 50 | 39.8 | 38.2 | 37.5 | 38.2 |
| 60 | 37.7 | 33 | 37.5 | 33 |
| 70 | 31.5 | 26.5 | 32.4 | 26.5 |
| 80 | 25.3 | 21.3 | 24.3 | 21.3 |
| 90 | 21 | 9.15 | 21.2 | 9.15 |
| 100 | 13.3 | 1.47 | 9.46 | 1.47 |

TABLE XXII

Value of bandwidth allocation for four FISs when access latency is kept constant and traffic priority is varied

| Input 13 Traffic Priority | Gaussian Mamdani FIS | Linear Sugeno FIS | Triangular Mamdani FIS | Constant Sugeno FIS |
|---|---|---|---|---|
| 10 | 31.6 | 28.8 | 28.3 | 28.8 |
| 20 | 33.1 | 31.1 | 31.1 | 31.1 |
| 30 | 35 | 33.5 | 33.4 | 33.5 |
| 40 | 37.3 | 35.8 | 35.5 | 35.8 |
| 50 | 39.8 | 38.2 | 37.5 | 38.2 |
| 60 | 42.5 | 40.6 | 39.5 | 40.6 |
| 70 | 44.9 | 42.9 | 41.6 | 42.9 |
| 80 | 47 | 45.3 | 44 | 45.3 |
| 90 | 48.7 | 47.6 | 46.7 | 47.6 |
| 100 | 50 | 50 | 50 | 50 |

In Table XXI, traffic priority is kept constant at 50 (moderate) and access latency is tuned in the range of 10 to 100 with a step size of 10, and their subsequent effects on the four FISs are calculated. It is noticed that as the value of access latency increases, the value of bandwidth allocation decreases for all the four FISs models.

In Table XXII, access latency is kept constant at 50 (moderate) and traffic priority is varied from 10 to 100 in the step size of 10, and its effect on the four FISs is evaluated. It can be observed that as the value of traffic priority increases, the value of bandwidth allocation increases for all the four models.

Moreover, a comparative study between Mamdani and Sugeno models are also carried out in Table XXIII, in order to calculate the correlation between two variables the mathematical formulae is used as given in [3]:

$$\text{Correlation}(x, y) = \frac{N \sum (x * y) - \sum x \sum y}{(\sum x^2 - (\sum x)^2)(\sum y^2 - (\sum y)^2)}$$

TABLE XXIII

Correlation between different FIS models

| Input Parameter | Gaussian and Triangular MF Mamdani | Constant and Linear MF Sugeno | Gaussian MF Mamdani and Linear MF Sugeno |
|---|---|---|---|
| Signal strength | 0.987783 | 1 | 0.996981 |
| Spectrum Demand | 0.998153 | 1 | 0.989348 |
| Signal To Noise Ratio (Signal to Interference plus Noise Ratio) for Channel Selection | 0.993698 | 1 | 0.992469 |
| Signal To Noise Ratio (Signal to Interference plus Noise Ratio) for Handoff | 0.998792 | 1 | 0.988249 |
| Interference | 0.998168 | 1 | 0.985673 |
| Channel Quality | 0.997753 | 1 | 0.974367 |
| Susceptibility | 0.997683 | 1 | 0.965043 |
| Spectrum Utilisation Efficiency | 0.99703 | 1 | 0.834503 |
| Degree of Mobility | 0.997396 | 1 | 0.911509 |
| Distance to Primary User | 0.997433 | 1 | 0.93538 |
| Secondary User Traffic Intensity | 0.982894 | 1 | 0.997091 |
| Bandwidth Allocation Traffic Intensity | 0.987463 | 1 | 0.995823 |
| Access Latency | 0.997389 | 0.992884 | 0.976432 |
| Traffic Priority | 0.990073 | 1 | 0.997219 |

Here x and y are the two values from respective models taken for comparison. $\sum x$ is the summation of variable x and $\sum y$ is the summation of variable y. Similarly, $\sum x^2$ is the summation of squared values of variable x and $\sum y^2$ is the summation of squared values of variable y. N is the number of data samples taken which are 10 in our case. The comparison between Mamdani and Sugeno model is performed in the present work in three different cases as given below:

1. Comparison between Gaussian Mamdani FIS and Triangular Mamdani FIS.
2. Comparison between Constant Sugeno FIS and Linear Sugeno FIS.
3. Comparison between Gaussian Mamdani FIS and Linear Sugeno FIS.

It is observed from Table XXIII that, it is clear that the for each input variable correlation value is very close to 1 in most cases, that means the respective models are highly correlated to each other. Therefore, the models can be used interchangeably.

## VI. CONCLUSION

To summarise, with this we can infer the following conclusion from the tables/figures presented above:

1. In Mamdani FIS, Gaussian MF as input gives closer expected value than Triangular MF, and it is clearly observable from Table IX–XXII. Furthermore, on the basis of above observation it may be concluded that Gaussian MF as input is better choice than Triangular MF as input.
2. In Sugeno FIS, Linear MF and Constant MF gives almost same value, which can be observed from Table IX–XXII. Considering the defuzzification process for both linear and constant MFs for all 13 inputs, it can be safely concluded that both Constant and Linear MF gives the same output value. Therefore, a Sugeno FIS will have very little effect when a particular MF is preferred for output over other MF (either Linear MF or Constant MF).
3. A high correlation (>0.95) is observed between Mamdani and Sugeno FIS, and it is evident that these FIS can be used interchangeably as it can be seen in Table XXIII.
4. Sugeno FIS is computationally more efficient than Mamdani for over 100 rules, as depicted in Table II. It is noticed that for smaller FIS (up to 30) rules, both FIS can be used interchangeably.
5. It is demonstrated through Table IX–XXII that Sugeno FIS is more accurate than Mamdani FIS, when we are varying the value of one parameter and keeping the other parameter values constant. In addition to that Sugeno FIS is more dynamic than Mamdani FIS to the input parameters, and offer more crisp output. However, the output behaviour of Mamdani FIS is more consistent than Sugeno FIS.
6. Gaussian MF has smoother curves for both line and surface plots than Triangular MF which is desirable in real life to maintain consistency.

It can be concluded based on the above discussed inferences, Mamdani and Sugeno FIS have high correlation, and in case of Mamdani FIS Gaussian MF as output is a better option than Triangular MF as output. Moreover, Sugeno FIS is a better option when computational time and accuracy is a factor and Mamdani FIS is a better option when consistency is a factor.

Therefore, as both the FIS have their own set of advantages and disadvantages so the selection between Sugeno FIS and Mamdani FIS depends on the application requirement.

## VII. FUTURE WORKS

There are various defuzzification techniques for Mamdani FIS, like

1. Centroid Method
2. Max membership principle: (Also known as the height method)
3. Max-membership principal
4. Weighted average method
5. Mean-max membership (Middle of maxima)
6. Center of sums
7. Center of Largest area
8. Largest of Maximum
9. Smallest of Maximum
10. Bisector Method

Out of which in the proposed work centroid method is used to evaluate crisp output. Furthermore, to implement the FIS in real world, it requires a careful approach in obtaining the real training data for training FIS and choosing the most appropriate defuzzification technique, as there is no generality in choosing defuzzification technique in different contexts and situations. Therefore, choosing a defuzzification technique will be possible only with real training data, and not with hypothetical data.

In addition to that, these different FISs can be joined to create a Fuzzy Neural Network, which can be further applied to implement the properties of Neural Networks for gaining more spectral utilisation and enhancing spectral efficiency

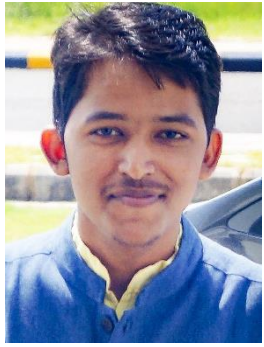

**Ashish Upadhyay** was born in Ghazipur, Uttar Pradesh, India in August 25 [th], 1998. He did his schooling from D.A.V. Public School, Moonidih, Dhanbad, Jharkhand, India and is currently pursuing Bachelors from Dr. Shyama Prasad Mukherjee International Institute of Information Technology, Naya Raipur (IIIT-NR), Chhattisgarh, India. He is co-founder and co-ordinator of public speaking society, Antardhwani at IIIT-NR. His main research interests are Artificial Intelligence, Cognitive Science, Signal Processing, Machine Learning and their applications.

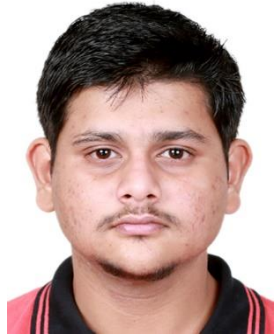

**Shashank Kotyan** was born in Indore, Madhya Pradesh, India in December 27 [th], 1997. He did his schooling from Holy Cross Senior Secondary School, Raipur, Chhattisgarh, India and is currently pursuing Bachelors from Dr. Shyama Prasad Mukherjee International Institute of Information Technology, Naya Raipur (IIIT-NR), Chhattisgarh, India. He is co-founder and co-ordinator of public speaking society, Antardhwani at IIIT-NR. His main research interests are Artificial Intelligence, Cognitive Science, Signal Processing, Data Science and their applications in IoT. He is also a student member of IEEE affiliated in 2017.

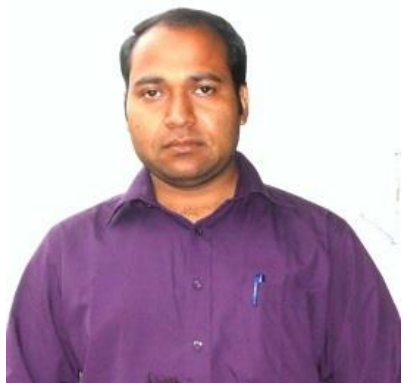

Shrivishal Tripathi is working as an assistant professor in International Institute of Information Technology, Naya Raipur. Prior to joining IIIT-NR, he was with BITS-Pilani, Hyderabad and NIIT University, Neemrana as an Assistant Professor. He has done his Ph.D. from Indian Institute of Technology Jodhpur in Electrical Engineering Department. He has received his M.E (Electronics and Electrical Communication) from PEC University of Technology, Chandigarh in 2011. He has published many papers in reputed Journals like IEEE Antenna and Wireless propagation Letters, Springer Wireless Personal Communication, IET Antenna and Propagation, IET Electronics Letters, Elsevier International Journal of Electronics and Communications, Journal of Electromagnetic Waves and Applications (Taylor and Francis), and Wiley's Microwave and Optical Technology Letters. Moreover, several papers presented in many IEEE international conferences (held in India and abroad).In addition to that he is an active reviewer of many reputed journals, and serve as TPC member for several reputed conferences organised in India and abroad. He reviewed many research fund proposal for various organisation/universities, too. He has organised many workshops in the areas of RF/Microwave and communication. He is a member of IEEE society and MTT society. His main research interest includes UWB Antenna, MIMO Antenna, Reconfigurable Antenna, SIW Antenna, On-body BAN Communication, SAR Analysis, Cognitive Radio and RF/Microwave circuit design.

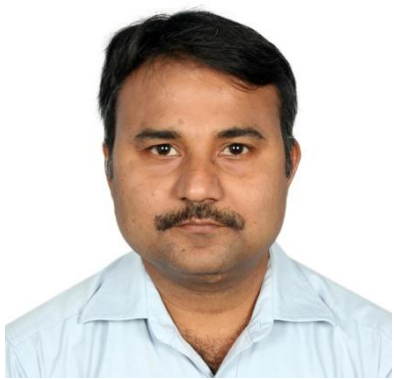

Sandeep Yadav is working as Assistant Professor in Indian Institute of Technology Jodhpur, Rajasthan, and received his PhD in Electrical Engineering from Indian Institute of Technology Kanpur in 2010. He is a recipient of SSI Young Scientist Award (2011), NI GSD Award (2012), and IBM shared Research University Award (2014). His research interests include capabilities and limits of Signal Processing, Condition Monitoring, Communication systems, Image Processing, Blind Source Separation, Artificial Neural Network and Healthcare applications. He has published many papers in reputed Journals like IEEE, IET, Wiley, Springer, etc.